\documentclass{article}
\usepackage[utf8]{inputenc}
\usepackage{amsmath}
\usepackage{amsthm}
\usepackage{amssymb}
\usepackage{mathabx}
\usepackage{xcolor}
\usepackage{bm}
\usepackage{graphicx}
\usepackage{bbm}
\usepackage{subcaption}
\usepackage{wasysym}
\usepackage{hyperref}
\usepackage[linesnumbered,ruled]{algorithm2e}
\usepackage{adjustbox}
\usepackage{mathbbol}
\usepackage{authblk}
\definecolor{lucacolor}{RGB}{0, 153, 51}

\newtheorem{lemma}{Lemma}
\newtheorem{theorem}{Theorem}

\theoremstyle{definition}
\newtheorem{definition}{Definition}
\newtheorem{remark}{Remark}

\allowdisplaybreaks[4]

\def \MMn{\mathcal{M}_n}
\def \Uij{U^{(j)}_i}
\def \Uj{\mathcal{U}^{(j)}}

\title{An Integer Linear Programming Model for Tilings}

\author[1]{Gennaro Auricchio \thanks{gennaro.auricchio@unipv.it}}
\author[1]{Luca Ferrarini \thanks{l.ferrarini3@campus.unimib.it}}
\author[1,2]{Greta Lanzarotto \thanks{g.lanzarotto@campus.unimib.it}}

\affil[1]{Department of Mathematics, University of Pavia}
\affil[2]{IRMA, University of Strasbourg}

\begin{document}

\maketitle

\abstract{

    \noindent In this paper, we propose an Integer Linear Model whose solutions are the aperiodic rhythms tiling with a given rhythm A. 
    
    \noindent We show how this model can be used to efficiently check the necessity of the Coven-Meyerowitz's $(T2)$ condition and also to define an iterative algorithm that finds all the possible tilings of the rhythm A. 
    
    \noindent To conclude, we run several experiments to validate the time efficiency of this model.
    % \noindent In this paper, we propose an Integer Linear Programming model for computing all the aperiodic tiling rhythmic canons given a generic rhythmic motif and a period. 
    % We characterize those rhythms as the solutions of an integer linear problem. 
    % %By introducing auxiliary binary variables, we impose the aperiodicity conditions as additional linear constraints. 
    % %To the best of our knowledge, this is the first time aperiodicity constraints are implemented. 
    % To conclude, we run several experiments to validate the efficiency of our model.
    }

\vspace{1cm}
% REQUIRED
\small{\textbf{Keywords}: Integer Programming, Mathematics and Music, Tiling Problems, Vuza Canons, $(T2)$ Conjecture
\medskip

\textbf{AMS}: 90C10, 05B45}

\section{Introduction}

In this paper, we deal with the mathematical and computational aspect of a musical problem that arouses the interest of mathematicians, computer scientists, music theorists, and composers (see \cite{Amiot2004} and \cite{Andreatta2004}).
It is about the construction of Vuza canons. 
A Vuza canon is a musical rhythmic canon without internal repetitions, regardless of the pitch, through which the composer tries to completely fill the rhythmic space, with no superimposition between the different voices \cite{Vuza}.\newline
\medskip

\noindent The construction of musical canons has always intrigued musicians: think of the complex Flemish polyphonies of composers such as Josquin Desprez or the counterpoint techniques that Johann Sebastian Bach shows in the \emph{Goldberg Variations}. The formal properties of the latter have been translated into algebraic terms in the work of Scimemi \cite{Scimemi}.
Olivier Messiaen is perhaps the first theorist and composer to have introduced and studied the concept of rhythmic canon regardless of pitch \cite{Messiaen}.\newline
\medskip

\noindent From a mathematical point of view, the construction of tiling rhythmic canons can be formalized in terms of factoring abelian groups as the sum of subsets.
Another representation makes use of polynomials with coefficients 0 and 1. 
It is in terms of these polynomials that the Coven-Meyerowitz conditions $(T1)$ and $(T2)$ are expressed, which are sufficient for the existence of rhythmic canons, and of which $(T1)$ is necessary (see \cite{Coven}).
The necessity of $(T2)$ remains an open problem and it is in this context that we find the central role of Vuza canons:

\begin{theorem}[Amiot, \cite{Amiot2005}]
If a rhythmic canon does not satisfy the $(T2)$ condition, it is possible to collapse it to a Vuza canon that does not satisfy the $(T2)$ condition.
\end{theorem}

\noindent Therefore, being able to compute Vuza canons and checking if a given rhythm tiles or not, has become a problem of major interest in the mathematical music field. \newline
\medskip

\noindent In this paper, we introduce a linear problem whose binary solutions are all the aperiodic tiling complements of a given rhythm. In particular, we impose the aperiodicity of the solution through linear constraints, at the best of our knowledge, this is the first time.
\medskip

\noindent The purpose of our model is twofold.

\noindent First, we want to determine, for a given motif $A$, all the tiling motifs $B$ in $\mathbb{Z}_{n}$. 
In this case, we are interested not only in testing the tiling property but also in finding all the complements of $A$.
Given a motif $A$ and a period $n$, the Matolcsi and Kolountzakis' \textit{Fill-Out Procedure}  provides a complete classification of the complements of $A$ in $\mathbb{Z}_n$ \cite{doi:10.1080/17459730903040899}. 
The main idea behind this algorithm is to use packing complements and add one by one the new elements discovered by an iterative search.
%heuristic search in an iterative way. 
At the best of our knowledge, this is the only algorithm able to provide the complete list of complements of a given motif, for $n\leq 200$.
For larger $n$ the problem has been considered in \cite{Jedrzejewski2013EnumerationOV}, but the author was able to give only a lower bound to the number of tiling complements. 
Therefore, we choose to compare our performances with the one of the \textit{Fill-Out Procedure}.

\noindent Secondly, we aim to determine if a given aperiodic motif $A$, that does not satisfies the ($T2$) property, tiles with an aperiodic motif $B$. 
This could be used to efficiently test possible counterexamples to the necessity of ($T2$) condition \cite{Amiot2011}.\newline
\medskip

\noindent The tiling problem is very similar to the decision problem of DIFF studied in \cite{Matolcsi}, which is shown to be NP-complete. 
This suggests a lower bound on the computational complexity of the tiling decision problem. Since our problem consists in solving a linear system of $3n-1$ unknowns and $3n+3(M_n(p)-1)$ constraints, the complexity of finding a single aperiodic solution is $O(n^c+3M_n(p))$, where $M_n(p)$ denotes the number of all distinct primes in the factorization of $n$.\newline
\medskip

%If we remove a founded solution from the set of feasible solutions and solve again the restricted problem, we will find a different solution. 
\noindent As we will see, solving this linear problem finds us only one of the possible solutions. However, we can update the problem by removing the founded solution from the feasible set. If we solve the updated problem, we are then able to find a new solution. By iterating this process until the problem cannot be solved, we will find all the tiling complements of the given rhythm $A$.

\noindent Since we are not interested in looking for all the possible solutions but rather for all the classes of equivalents rhythms modulo translations or affine transformations, we can costumize the constraints to add at each step. 
In particular, if we are interested in finding all the solutions modulo affine transformations, the number of constraints to add at each iteration it is equal to the cardinality of $\mathcal{P}=\{a \in \mathbb{N} \mid (a,n)=1 \}$ times the cardinality of the set of all translations fixing the first entry of the solution equal to $1$.
Therefore, we add $O(|\mathcal{P}|\frac{n}{n_A})$ new constraints at every iteration, where $n_A$ is the cardinality of the rhythm $A$.
As a result, finding new tiling rhythms gets harder at each iteration. \newline
\medskip
% We refer to this effect as the tail effect.

%%%%%%%%%%%%%%%%%%%%%%%%%%%%%%

\noindent The outline of the paper is the following.

\noindent In Section \ref{sec:basic_notions}, we recall the main notions and results about Tiling Rhythm Canons and formulate the tiling problem.

\noindent In Section \ref{sec:tiling_model}, we reformulate the tiling problem as an Integer Linear Problem. We endow the obtained system with additional constraints to impose the aperiodicity of the solution. We then define an iterative algorithm able to compute the complete tiling of a given rhythm.

\noindent In Section \ref{sec:num_res}, we report the results of our tests. We compare the time required by our method with the one required by the \emph{Fill-Out Procedure}.
%one of the most used methods nowadays. %Moreover, we are able to improve the lower bounds on the cardinality of the possible tiling cannons, presented in \cite{Jedrzejewski2013EnumerationOV}.

\noindent To conclude, in Section \ref{conclusion}, we outline the future works and possible research directions.

\section{Tiling in music}
\label{sec:basic_notions}
In this section, we fix our notation and recall the main notions about rhythm in mathematics. We refer to \cite{Amiot2011} for a complete and exaustive tractation of this topic.
%of those topics.
    \begin{definition}\label{df:TRC}
		A \emph{tiling rhythmic canon} (\emph{TRC}) $(A,B)$ with \emph{period} $ n $ is a factorization of the cyclic group $\mathbb{Z}_{n}$ given by subsets $A$, the \emph{inner rhythm}, and $B$, the \emph{outer rhythm}:
		\[A\oplus B=\mathbb{Z}_{n}.\]
	\end{definition}
	
	\noindent Fixed $n\in\mathbb{N}$, a classical problem is to determine if, given an \emph{inner rhythm} $A$, there exists an \emph{outer rhythm} $B$.	It is possible to characterize TRCs through characteristic polynomials.
	
	\begin{definition}
	    Let $A\subset{\mathbb{N}}$ be finite. 
	    The \emph{characteristic polynomial} of $A$ is defined as
	    \[
	    p_A(x)=\sum_{k\in A}x^{k}.
	    \] 
	\end{definition}

    \begin{lemma}
    \label{lm:pol_equivalence}
        Let $p_A (x), p_B (x) \in \mathbb{N} [x]$ and $n$ a positive integer. Then
        \[
        p_A (x)\cdot p_B (x)\equiv \Delta_{n} (x),\quad\quad\mod (x^{n} - 1)
        \]
        if and only if
            \begin{enumerate}
                \item $p_A (x), p_B (x) \in \{0, 1\} [x]$, and
                \item $A\oplus B = \{r_{1}, \dots, r_{n}\} \subset\mathbb{Z}$ with $r_{i}\neq r_{j} \mod (n)$ for each $i, j \in \{1,\dots, n\}, i\neq j$.
            \end{enumerate}
        \end{lemma}

	\begin{remark}
	Note that
	\[
	\Delta_{n}(x)=\frac{x^{n}-1}{x-1}=\prod_{\substack{ d\divides n\\d\neq1} }\Phi_{d}(x)
	\]
	where $\Phi_{d}(x)$ is the $d-$th \emph{cyclotomic polynomial}, that is the minimal polynomial of any primitive $d-$th root of unity over the field of the rational numbers.
	\end{remark}

    \noindent An important property exploited in our algorithm is the invariance of solutions under affine transformations, that is, any affine transformation sends tiling solutions into tiling solutions.
    
    \begin{theorem}[Vuza, \cite{Vuza}]
    \label{Vuza}
        Let $A\oplus B=\mathbb{Z}_{n}$ be a TRC and $f:\mathbb{Z}_n \to \mathbb{Z}_n$  be an affine transformation of $\mathbb{Z}_n$, that is
        \[
        f:x\mapsto ax+b\quad\quad \mod n,
        \]
        where $a$ is coprime with $n$ and $b\in \mathbb{Z}_n$. The affine transform of $A$ by $f$ still tiles with $B$; i.e. $(aA+b)\oplus B=\mathbb{Z}_{n}$.
    \end{theorem}
    
    \begin{definition}
        Let $k$ be a non-null element of $\mathbb{Z}_{n}$. A rhythm $A\subset \mathbb{Z}_{n}$ is \emph{periodic modulo} $k$ if and only if  $k + A = A$. A rhythm $A\subset \mathbb{Z}_{n}$ is \emph{aperiodic} if and only if it is not periodic for any $k \in \mathbb{Z}$.
    \end{definition}
    \begin{remark}\label{aperiodic}
    Note that a set $A$ is periodic modulo $k\divides n$ if and only if
    \[
    \frac{x^{n}-1}{x^k-1}\mathrel{\Big|} p_A(x).
    \]
    Whenever a rhythm $A$ is periodic modulo $k\divides n$, with $k\neq n$, it is periodic modulo all multiples of $k$ dividing $n$. For this reason, when it comes to check whether $A$ is periodic or not, it suffices to check if it is periodic modulo $m_1= p_1^{\alpha_1-1} p_2^{\alpha_2} \dots p_N^{\alpha_N}$, $m_2= p_1^{\alpha_1} p_2^{\alpha_2-1} \dots p_N^{\alpha_N}$, $\dots$, $m_N= p_1^{\alpha_1} p_2^{\alpha_2} \dots p_N^{\alpha_N-1}$, 
     where $n = p_1^{\alpha_1} p_2^{\alpha_2} \dots p_N^{\alpha_N}$ 
    %  %and each $p_j$ is a prime number. 
    is the prime powers
    %unique
    factorization of $n$.
    %into prime powers.
    \end{remark}

    \begin{definition}
    A TRC $(A,B)$ in $\mathbb{Z}_{n}=A\oplus B$ is a \emph{Vuza canon} if both $A$ and $B$ are aperiodic.
       % A \emph{Vuza canon} is a TRC $\mathbb{Z}_{n}=A\oplus B$ where neither $A$ nor $B$ is periodic.
    \end{definition}
    \noindent The existence of Vuza canons depends on the order of the cyclic group $\mathbb{Z}_{n}$ factorized. In \cite{Hajos}, Haj\'{o}s proposed the following definition.
    \begin{definition}
    A finite abelian group $G$ is a \emph{good group} if in any tiling $G = A \oplus B$ one of the two subsets $A$ and $B$ has to be periodic.  $G$ is a \emph{bad group} if there exists a tiling $G = A \oplus B$ where $A$ and $B$ are aperiodic.
    \end{definition}

    \noindent In \cite{Bru}, \cite{Hajos}, \cite{Redei}, and \cite{Sands} the good groups and the bad groups have been completely characterized. Moreover, they partition the set of finite cyclic groups in two disjoint classes. In particular: 
    
    % the class of finite cyclic groups is divided into two disjoint subclasses, explicitly defined:
    \begin{itemize}
        \item the good groups, for which there are no Vuza canons, have orders in
        \[
        \big\{p^{\alpha}, p^{\alpha}q, p^{2}q^{2}, pqr, p^{2}qr, pqrs:\alpha \in \mathbb{N}\big\},
        \]
        where $p, q, r, s$ are distinct primes, and
        \item the bad groups, whose orders are of the type $N = nmk$ with
    \begin{itemize}
        \item $(n,m) = 1$
        \item $n = n_{1}n_{2}$, $\;m = m_{1}m_{2}$
        \item $n_{1}, n_{2},\; m_{1}, m_{2},\; k \geq 2$.
    \end{itemize} 
    \end{itemize}
        Therefore, the analysis on Vuza canons  exclusively concern these last cyclic groups, whose orders are explicitly identified. %In particular, we focus on [72,108,120,144,168,180, $\dots$].
    Although the groups that can be expressed as direct sum decomposition have been identified exactly, it does not mean that every rhythm in $\mathbb{Z}_{n} $ tiles. 
    Ethan Coven and Aaron Meyerowitz found two sufficient conditions for a rhythmic pattern to tile \cite{Coven}.  
    Those conditions have been proved to be also necessary under certain hypothesis, however a proof of their necessity in the general case is still lacking. 
    The polynomial representation of TRCs turns out to be the most suitable for presenting these results.\newline
    \medskip
    
    \noindent To state the condition introduced by Coven and Meyerowitz, we need to define two sets on the basis of the cyclotomic polynomials which divide the characteristic polynomial of the rhythm under consideration.
    
    \begin{definition}
        Let $A \subset\mathbb{N}$ be finite, we define:
        \begin{itemize}
            \item $R_{A}: = \left\{ d \in \mathbb{N}^{*}: \Phi_{d}(x)\divides p_A (x) \right\}$,
            \item $S_{A}: = \left\{ d \in R_{A}: d = p^{\alpha}, p \mbox{ prime},\alpha\in \mathbb{N}^{*}\right\}$,
        \end{itemize}
        where $\mathbb{N}^*:=\mathbb{N}\backslash \{0\}$.
    \end{definition}
    \noindent We can now state the following:
    \begin{theorem}[Coven and Meyerowitz, \cite{Coven}]
        Let us consider the conditions:
        \begin{itemize}
            \item[$ (T1) $] $A (1) =\prod_{p^{\alpha}\in S_{A}} p$;
            \item[$ (T2) $] if $ p_{1}^{\alpha_{1}},\dots,p_{N}^{\alpha_{N}}\in S_{A}$, then $ p_{1}^{\alpha_{1}}\cdots p_{N}^{\alpha_{N}}\in R_{A} $,
	where $ p_{1}^{\alpha_{1}},\dots,p_{N}^{\alpha_{N}} $ are powers of distinct primes.
        \end{itemize}
        Then
        \begin{enumerate}
        \item if A satisfies (T1) and (T2), then it tiles;
        \item if A tiles, then it satisfies (T1);
        \item if A tiles and $|A|$ has at most two prime factors, then A satisfies (T2).
        \end{enumerate}
    \end{theorem}
    \noindent Determining whether the condition $(T2)$ is necessary for a rhythm $A$ to tile is still an open question. 
%    A classical result states that $(T2)$ is necessary when the cardinality of the set $A$ is equal to the product of at most two prime powers.  However,  
Izabela \L{}aba and Itay Londner were able to prove that the condition $(T2)$ holds for all integer tilings of period $M = (p_ip_jp_k)^2$, where $p_i, p_j , p_k$ are distinct odd primes:

%\begin{theorem}[\L{}aba, \cite{Laba}]\label{Laba}
%		Let $ p,q,r $ be distinct primes. Assume $ A\oplus B=\mathbb{Z}_{n} $, $ |A|=p^{\alpha}q^{\beta}r^{\gamma} $, and $ |B|=pqr $. If $ \Phi_{p}(x) $, $ \Phi_{q}(x) $, and $ \Phi_{r}(x) $ divide $ p_A(x) $, then so do $ \Phi_{pq}(x) $, $ \Phi_{pr}(x) $, $ \Phi_{qr}(x) $, and $ \Phi_{pqr}(x) $.
%\end{theorem}
\begin{theorem}[\L{}aba and Londner, \cite{Laba21}]\label{Laba21}
		Let $M = p^{2}_{i}p^{2}_{j}p^{2}_{k}$, where $p_{i}, p_{j}, p_{k}$ are distinct odd primes. Assume that
$A\oplus  B = \mathbb{Z}_{M}$, with $|A| = |B| = p_{i}p_{j}p_{k}$. Then both $A$ and $B$ satisfy (T2).
\end{theorem}

\section{A Linear Model for tiling in \texorpdfstring{$\mathbb{Z}_{n}$}{TEXT} }
\label{sec:tiling_model}

In this section, we introduce our Integer Linear model. First of all, we define the linear equations that describe the tiling property. Afterwards, we impose the aperiodicity constraints. Our main result is Theorem \ref{thm:aperiodic_solutions}, where we state that imposing the aperiodicity of the solution can be done through linear constraints. Finally, we show how solving a sequence of increasingly harder linear problems leads to a complete tiling of a given rhythm $A$.

\subsubsection*{Feasibility Condition}
	Let us take an inner rhythm $A$ and a possible outer rhythm $B$. Since the degrees of their characteristic polynomials, $p_{A}(x)$ and $p_{B}(x)$, are both less than or equal to $n-1$, the degree of the product $p_R(x)$ is less than or equal to $2n-2$. We denote by $r$ the vector with $2n-1$ entries containing the coefficients of the polynomial $ p_{R}(x):=p_{A}(x)p_{B}(x) $.
	From Lemma \ref{lm:pol_equivalence}, we know that $B$ tiles with $A$ if and only if
	\begin{equation}
	\label{eq:equiv_p_R}
	    	p_{R}(x)\equiv 1+x+x^2+\dots+x^{n-1}, \quad\quad\quad\mod x^{n}-1.
	\end{equation}
	We can express condition \eqref{eq:equiv_p_R} through $n$ linear equations 
	\[ 
	r_{i}+r_{i+n}=1,\quad\quad\quad\forall i=0,\dots, n-1. 
	\]
% 	These relations alone imply that
% 	\[ 
% 	\sum_{i=0}^{2n-1}r_{i}=n. 
% 	\]
	Therefore, we can express the constraint 
	\[
	p_{R}(x)=p_A(x)p_B(x)\equiv\sum_{i=0}^{n-1}x^{i},\quad\quad \mod x^{n}-1,
	\]
	through the linear system
    \[
	    \begin{split} 
		F_{i}(B)-r_{i}&=0\quad\quad\quad\forall i\in\{0,\dots,2n-2\},\\
		r_{j}+r_{j+n}&=1\quad\quad\quad\forall j\in\{0,\dots,n-1\},
	\end{split}
    \]	
    
    \noindent where $F_{i}(B)$ is the function that associates to a motif $B$ the $i$-th coefficient of $ p_A(x) p_B(x) $, that is
	\[ 
	\begin{split}
		F_{0}(B)&:=a_{0}b_{0},\\
		F_{1}(B)&:=a_{1}b_{0}+a_{0}b_{1},\\
		F_{2}(B)&:=a_{2}b_{0}+a_{1}b_{1}+a_{0}b_{2},\\
		\vdots&\qquad\vdots\\
		F_{2n-2}(B)&:=a_{n-1}b_{n-1},
	\end{split} 
	\]
	where $b_0,b_1,\dots,b_{n-1}$ are the coefficients of $p_B$. Notice that, since $A$ is given, all the equations presented above are linear with respect to the variables $b_i$ and $r_i$. We then can express them through a linear system
	\begin{equation}
	\label{eq:tiling linearsystem}
	\mathcal{A}\cdot\mathcal{X}=\mathcal{Y}, 
	\end{equation}
	where \begin{itemize}
		\item $ \mathcal{A} $ is a $ (3n - 1) \times(3n - 1) $ matrix which depends only on the given rhythm $ A $,
		\item$ \mathcal{X} = (b,r) $ is the vector composed by  the coefficients of $p_B$ (namely $b$) and the coefficients $p_R$ (namely $r$), respectively;
		\item$ \mathcal{Y} $ is the $ (3n-1) $-dimensional vector defined as
		\[
		\mathcal{Y}_i=\begin{cases}
		0 \quad \quad \text{if}\; i\in \{0,\dots,2n-2\},\\
		1 \quad \quad \text{otherwise.}
		\end{cases}
		\]
	\end{itemize}
Finally, in order to ensure that $p_B$ and $p_R$ are $0-1$ polynomials, we will require $b_i$ and $r_i$ to be binary variables, i.e. they can only assume value $0$ or $1$.

\subsection*{Aperiodicy Constraints}

Let us assume $n=p_1^{\alpha_1}p_2^{\alpha_2} \dots p_N^{\alpha_N}$. Without loss of generality, we can suppose
\[
p_1<p_2<\dots<p_N
\]
and, therefore, if we define the set of the maximal divisors of $n$ as $\mathcal{M}_n:=\{m_k=\frac{n}{p_k}\}_{k=1,\dots,N}$, we have
\[
m_N<m_{N-1}<\dots<m_1.
\]
According to Remark \ref{aperiodic}, to verify if the rhythm $B$ is periodic or not, it is sufficient to check its periodicity only for the elements of $\MMn$.

\noindent Let us take $m_j\in\MMn$. To impose that the rhythm $B$ is not $m_j-$periodic, we introduce the family of auxiliary variables 
\[
\Uj:=\big\{\Uij\}_{i=0,\dots,m_j-1}.
\]
Each family $\Uj$ is composed by binary variables subjected to the the following constraints:
\begin{align}
\label{eq:cond_aper1}&\sum_{k=0}^{p_j-1}b_{i+km_j}-p_j\Uij\leq p_j-1,\\
\label{eq:cond_aper2}&\sum_{k=0}^{p_j-1}b_{i+km_j}-p_j\Uij\geq 0,\\
\label{eq:cond_aper3}&\sum_{i=0}^{m_j-1}\Uij\leq \frac{n_B}{p_j}-1,
\end{align}
for each $j$ such that $p_j | n_B$, where $n_B$ is the cardinality of $B$.

\noindent Since $\sum_{k=0}^{p_j-1}b_{i+km_j}\leq p_j$, condition \eqref{eq:cond_aper1} assures us that $\Uij=1$ if 
\[
\sum_{k=0}^{p_j-1}b_{i+km_j}=p_j.
\]
Condition \eqref{eq:cond_aper2} assure us that $\Uij =1$  only if 
\[
\sum_{k=0}^{p_j-1}b_{i+km_j}=p_j.
\]
Therefore, conditions \eqref{eq:cond_aper1} and \eqref{eq:cond_aper2} combined, assures us that
\[
\Uij=1 \iff \sum_{k=0}^{p_j-1}b_{i+km_j}=p_j.
\]
Since $\sum_{i=0}^{n-1}b_i=n_B$, if $\sum_{i=0}^{m_j-1}\Uij=\frac{n_B}{p_j}$, it follows that
\[
\sum_{k=0}^{p_j-1}b_{i+km_j}=\begin{cases}
p_j \quad \quad \text{if}\;\;\; \Uij=1\\
\;\\
0\quad \quad \text{otherwise,}
\end{cases}
\]
and, hence, $B$ is periodic of period $m_j$. By adding the constraints \eqref{eq:cond_aper1}, \eqref{eq:cond_aper2}, and \eqref{eq:cond_aper3} to the Linear System, we, therefore, remove all the periodic solutions from the feasible set.
%Condition \eqref{eq:cond_aper3} prevents this from happening.

\begin{remark}
To improve the efficiency, we remove a family of auxiliary variables $\Uj:=\big\{\Uij\big\}$ by imposing
\begin{equation}
    \label{eq:diff_inequality}
    \sum_{i=0}^{m_j-1}b_i\leq \frac{n_bm_j}{n}-1.
\end{equation}
Indeed, if $B=(b_0,b_1,\dots,b_{n-1})$ is not $m_j-$periodic, there exists a translation of $B$ such that \eqref{eq:diff_inequality} holds. We remove the family $U^{(0)}$, since it contains the highest number of variables.
\end{remark} 

\noindent Since conditions \eqref{eq:cond_aper1}--\eqref{eq:cond_aper3} and \eqref{eq:diff_inequality} are linear for any $j$, we can add them to the system described in \eqref{eq:tiling linearsystem} and obtain the following Integer Linear Programming (ILP) problem

\begin{align}
\label{m1:obj}  \min \quad & \mathcal{O}(\{b_i\},\{r_i\},U) \\
\label{m1:c1} \mbox{s.t.} \quad  & \sum_{j=0}^{i} a_{i-j}b_j - r_i = 0 & \forall i\in\{0,\dots,n-1\},\\
\label{m1:c2} & \sum_{j=0}^{i+1} a_{n-(i-j)}b_j - r_{i+n} = 0 & \forall i\in\{0,\dots,n-2\},\\
\label{m1:c3}  &  r_{j}+r_{j+n} =1 & \forall j\in\{0,\dots,n-1\}, \\
\label{m1:c4}  & \sum_{j = 0}^{m_0-1}b_{j} \leq n_B\frac{m_0}{n} -1  & \;\\
\label{m1:c5} & \sum_{k=0}^{p_j-1}b_{i+km_j}-p_j\Uij\leq p_j-1, & \forall j \in \{1,\dots,N\}\\
\nonumber & \quad & \forall i \in \{0,\dots,m_j-1\} \\
\label{m1:c6}&\sum_{k=0}^{p_j-1}b_{i+km_j}-p_j\Uij\geq 0,  & \forall j \in \{1,\dots,N\}\\
\nonumber & \quad & \forall i \in \{0,\dots,m_j-1\} \\
\label{m1:c7}&\sum_{i=0}^{m_j-1}\Uij\leq \frac{n_B}{p_j}-1, & \forall j \in \{1,\dots,N \}\\
\nonumber & \quad & \forall i \in \{0,\dots,m_j-1\} \\
\label{eq:b_zero} & b_0 = 1 \\
\nonumber & b_k \in \{0,1\} & \forall k \in \{1,\dots,n-1\},\\
\nonumber & r_k \in \{0,1\} & \forall k \in \{0,\dots,2n-2\},\\
\nonumber & \Uij \in \{0,1\} &  \forall j \in \{1,\dots,N\}\\
\nonumber & \quad & \forall i \in \{0,\dots,m_j-1\} \\
\nonumber
%\text{for}\; j=1,\dots,N\\
%\nonumber&\quad &\text{for}\; i=0,\dots,m_k-1.
\end{align}

\noindent where $\mathcal{O}$ is a suitable linear function to minimize. The constraint \eqref{eq:b_zero} allows us to reduce the size of the feasible set by removing a degree of freedom from the possible solution.  We denote the model just introduced as the Master Problem (MP). 

\begin{remark}
The set of constraints of the MP fully characterize the possible aperiodic rhythms tiling with a given rhythm $A$. The functional $\mathcal{O}$ does not play any role, however it can be used to induce an order or a selection criteria on the space of solutions. For example, let us consider the following functional
\[
\mathcal{O}(b,r):=\sum_{i=0}^{2n-2}i^2b_i.
\]
This functional prefers the tiling complements whose first components are as full as possible of ones. Choosing the right functional $\mathcal{O}$ can help in discerning, among all the possible solutions, the ones we want to find. However, since the aim of our tests is to find all the possible tilings, we will not need to impose any selection criteria and, therefore, we will set
\[
\mathcal{O}(b,r):=0
\]
for all the experiments.
\end{remark}

\begin{theorem}
\label{thm:aperiodic_solutions}
Given an inner rhythm $A$ in $\mathbb{Z}$, let $\hat{\mathcal{Y}}=(b,r)$  be a solution of $MP$. Then, the rhythm associated to the characteristic polynomial
\[
p_B(x):=\sum_{i=0}^{n-1}b_ix^{i},
\]
is aperiodic and tiles with $A$.
\end{theorem}

\noindent To find all the aperiodic complements in $\mathbb{Z}_n$ of a given rhythm $A$, is therefore equivalent to find all the solutions of the $MP$, such that
\[ 
\sum_{i=0}^{n-1}b_{i}=n_B:=\frac{n}{n_A}. 
\]
\noindent We denote with $D_A$ the set containing all these solutions.

\subsection*{Cutting Sequential Algorithm}
%
% Following \genna{(citare i lavori in cui fanno cose simili)} we also want to list all the possible rhythms using a cutting process.
%

\noindent Once we find an aperiodic rhythm $B^{(1)}$ tiling with a given rhythm $A$, we can remove $B^{(1)}$ from the set of all possible solutions $D_A$ and obtain a new set of feasible solutions $D^{(1)}_A$. Let us denote with $MP^{(1)}$ the restriction on $D_A^{(1)}$ of $MP$ and call $B^{(2)}$ the solution of $MP^{(1)}$, we can then remove this solution from $D^{(1)}_A$, define the set $D^{(2)}_A$, and define $MP^{(2)}$, starting the whole process again. 
By repeating this process until we find an unsolvable problem, we retrieve all the possible solutions of the original Master Problem and, therefore, we generate all the aperiodic rhythms tiling with the rhythm $A$.\newline
\medskip

\noindent In this paragraph, we detail how to cut out from the feasible set the solution found at each iteration.\newline
\medskip

\noindent Let $B^{(1)}$ be a rhythm tiling with $A$ and let $ b^{(1)} = (b_{0},\dots,b_{n-1})$ be the coefficients of its characteristic polynomial. We denote with $ I^{(1)} $ the set of non-zero coordinate indexes of the vector $ b^{(1)} $, that is
\[ 
 I^{(1)}:=\Big\{i\in\{0,\dots,n-1\}\;\Big|\; b_{i}=1\Big\}.
\]
If we add the constraint 
\begin{equation}
\sum_{i\in I^{(1)}}b^{(1)}_{i}\neq\frac{n}{n_A}, 
\end{equation}
or equivalently 
% as a linear constraint
\begin{equation}\label{iterate}
\sum_{i \in I^{(1)}}b^{(1)}_{i}\leq\frac{n}{n_A}-1,
\end{equation}
to the MP %\eqref{eq:tiling linearsystem} 
and solve it, we find a new solution $b^{(2)}\neq b^{(1)}$ of the tiling problem. We iterate this procedure until we find an unsolvable problem. All the solutions found during this process are stored in memory and given as final output of the algorithm.

\noindent In Algorithm \ref{alg_CSA}, we sketch the pseudocode of this algorithm.
\begin{algorithm}[t!]
    \caption{ The Cutting Sequential Algorithm. \label{alg_CSA}}
    \SetKwInOut{Input}{Input}
    \SetKwInOut{Output}{Output}

    \Input{rhythm $A$}
    \Output{$S$, list of Aperiodic rhythms $B$, such that $A \oplus B = \mathbb{Z}_n$}
    $z^{*}= OPT(MP)$
    \\ add $z^{*}$ to $S$
    \\ \While{$P\neq \emptyset$}
    { add $\sum_{i \in I_z} b_i \leq \beta$ to ($MP^{(i)}$)
    \\ Solve ($MP^{(i)}$)
    \\ $z_{new}$ = OPT($MP^{(i)}$)
    \\ set $I_{z}:=I_{z_{new}}$
    \\ add $z_{new}$ to S
    }
    return S
\end{algorithm}

\begin{remark}
Adding the constraints one by one is highly inefficient. Therefore, once we find a solution, we compute all its affine transformations, which, according to Theorem \ref{Vuza}, are possible solutions and remove them as well. Since we impose $b_0=1$, we consider only the affine transformations that preserve this identity.
This procedure, however, is customizable: if we remove only the translations of the founded solution the algorithm will return all the solutions modulo translations.
\end{remark}

\noindent Given a solution $b^{(1)}$, we can remove the affine transformations of a given solution through a linear constraint. According to \eqref{iterate}, we impose
\begin{equation}
    \sum\limits_{i \in I^{(1)}} b_{a(i+k)} \leq n_B-1
\end{equation}
where $k$ runs over all the translations which fix the first position and $a$ runs over the set of numbers co-prime with $n$.

\subsubsection*{Complexity of the Method}

To conclude, we analyze the complexity of the system \eqref{eq:tiling linearsystem}. The unknowns to determine are the $3n-1$ binary coordinates of the vector $(b,r)$ plus the variables needed to impose the aperiodicity constraints, $\Uij$, which are 
\[
\sigma_n:=\sum_{p\in \mathbb{P}_n\backslash\{p_0\}}\frac{n}{p},
\]
where $\mathbb{P}_n$ is the set of primes that divide $n$. Therefore, we have $3n-1$ constraints for the feasibility, the $3\sigma_n$ given by conditions \eqref{m1:c5}, \eqref{m1:c6}, and \eqref{m1:c7} plus the one given by condition \eqref{m1:c4}. Since it is well-known that 
\[
\lim_{n\to \infty}\sum_{p\in\mathbb{P},p\leq n}\dfrac{1}{p}\approx \lim_{n\to \infty}\log(\log(n))=+\infty,
\]
it is impossible to give a bound on the number of aperiodicity constraints that does not depend from $n$.

\noindent If we want a complete tiling of the given rhythm, the complexity increases, since we are adding constraints at each iteration. The amount of constraints to add depends on the equivalence class we are computing. If we are looking for all the solutions modulo translation, we add $n_B$ constraints at each iteration, since there are exactly $n_B$ feasible translations preserving the identity $b_0=1$. If we search for all the solutions up to affine transformations, the number of constraints added is $n_B$ times the quantity of numbers primes to $n$.

\section{Numerical Results}
\label{sec:num_res}

In this section, we report the results of our tests. Our experiments aim in showcasing the efficiency and the quickness of our model. We inhabit our tests in two frameworks. 

\noindent In the first one, we aim to find all the complements of a given rhythm. We compare the CSA with the \emph{Fill-Out Procedure} on rhythms in $\mathbb{Z}_n$, for $n=72,108,120,144,168,180$. In the second one, we want to determine if a given rhythm tiles with an aperiodic rhythm, i.e. we want to find just one of the possible complements of a given rhythm. This simplification allows us to test our methods on larger values of $n$.\newline
\medskip

\noindent We run all our experiments on a ASUS VivoBook15 with Intelcore i7. The algorithm is implemented in Python using Gurobi v9.1.1, \cite{gurobi}.

\subsection{Runtimes for Complete Tilings}
We tested our method and the \emph{Fill-Out Procedure} on several rhythms in various $\mathbb{Z}_n$, for $n=72,108,120,144,168,180$. 
% It is worth of mention that the \emph{Fill-Out-Procedure} finds every complement modulo translation, while the CSA can be costumized to compute every complement modulo all the affine transformations. 
% We choose to run our method to compute all the solutions modulo affine transformation,
The experiment we ran is the following. Given a rhythm $A$, we list every complement. Afterwards, we reverse the problem: we fix one of the found complements, namely $B$, and search for all the complements of $B$.\newline
\medskip

\noindent In Table \ref{table_times}, we compare the runtimes of CSA with the runtimes of the \emph{Fill-Out Procedure}. The CSA is customized in order to find all the classes modulo affine transformations.

\subsubsection*{The Tail Effect}
\label{sec:Tail_Effect}

% As we stated above, our method well behaves when there are few complements to find. 
Every time we find a solution, we have to add new constraints to the Master Problem and solve it once again. As a result, the problem we solve gets computationally harder at each iteration. In particular, the time needed to compute the last complements of a given rhythm requires way more time than computing the first half.\newline
\medskip

\begin{figure}[ht]
    \centering
    \begin{minipage}[b]{1\textwidth}
        \centering
        \includegraphics[width=0.75\linewidth]{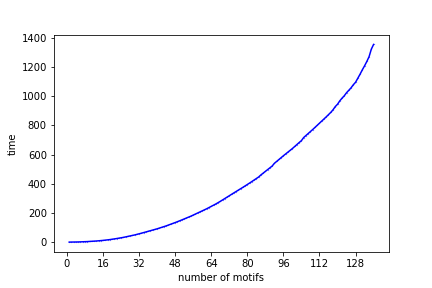}
       % \small\caption{$n = 180,A=\{0, 12, 24, 45, 57, 69\}$}
        % \label{fig:136}
    \end{minipage}\\
    
    \begin{minipage}[b]{1\textwidth}
        \centering
        \includegraphics[width=0.75\linewidth]{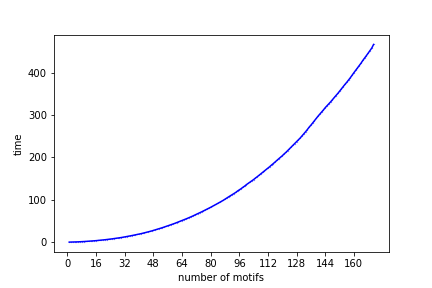}
        %\small\caption{$n = 180,A=\{0, 12, 24, 36, 45, 48, 57, 69, 81, 93\}$}
    \small{\caption
    {
    Time (in seconds) to find the next solution with CSA for two rhythms in $\mathbb{Z}_{180}$. 
    On the top $A=\{0, 12, 24, 45, 57, 69\}$, on the bottom $A=\{0, 12, 24, 36, 45, 48, 57, 69, 81, 93\}$. \label{fig:171}
    }}
    \end{minipage}
    
\end{figure}

\noindent In Figure \ref{fig:171}, we report the time required to find the next tiling solution for two rhythms in $\mathbb{Z}_{180}$. As expected, the time required at each iteration grows exponentially.

\subsection{Verifying the Tiling Property}

We are now interested in determine if a given rhythm $A$ admit an aperiodic tiling complement $B$. We believe that, by pairing our model with a function that builds a non $(T2)$ rhythm $A$, we could create a counter example to the necessity of this condition. For this reason, being able to verify the tiling property of a rhythm $A$ in a reasonable amount of time is important.\newline
\medskip

\begin{table}[t!]
    \centering

		\begin{tabular}{|| c | c  ||} 
			\hline
			$ n $ & Rhythm tested \\ 
            \hline
            $1050$ & $\{0, 15, 30, 35, 45, 60, 70, 75, 90, 105 \}$\\
            \hline
            % & 42.25\\\hline
            $2310$ & $\{0, 5, 6, 10, 12, 18, 24, 26, 30, 31, 36\}$  \\
            % 6.01
            \hline
            $6300$ & $\{0, 2, 4, 5, 6, 7, 8, 10, 12, 350, 352, 354, 355, 356, 357, 358, 360, 362 \}$ \\
            % & 42.25\\
            \hline
            
            $27225$ & $\{ 0, 9, 15, 18, 24, 27, 30, 36, 39, 45, 54, 3025, 3034, 3040,$\\
             \; & $3043, 3049, 3052, 3055, 3061, 3064, 3070, 3079, 6050, 6059,$ \\
             \; &    $6065, 6068, 6074, 6077, 6080, 6086, 6089, 6095, 6104 \}$\\  
        %  726.48\\
            \hline

	\end{tabular}
\small{\caption{Non $(T2)$ candidate rhythms checked.\label{enumvuza2}}}	
\end{table}

\noindent In Table \ref{enumvuza2}, we report the rhythms tested with our method. The runtimes required to determine the non-existence of an aperiodic complement varies in a range of $1$ minute (for the rhythms in $\mathbb{Z}_{1050},\mathbb{Z}_{2310}$, and $\mathbb{Z}_{6300}$) and up to $10$ minutes (for the rhythm in $\mathbb{Z}_{27225}$).

\section{Conclusions and Future Works}
\label{conclusion}
We introduced a new Integer Linear Model able to find the aperiodic complements of a given rhythm. We run several tests to prove the time efficiency of our method, especially when it comes to determining if there exists an aperiodic complementary of a given rhythm.\newline
\medskip

\noindent Our future aim is to characterize the polynomial induced by a rhythm that does not satisfy the $(T2)$ condition through a Linear Programming Model. This could lead to discovering insightful information on the structure of those canons. Moreover, by pairing an algorithm that quickly searches for non $(T2)$ motifs with the algorithm introduced in this paper, we hope to find a counterexample to the necessity of $(T2)$.\newline
\medskip

\noindent We also want to improve our algorithm further by dividing the set of solutions into smaller and disjoint sets. Hopefully, this division will mitigate the ``tail effect" showcased in subsection \ref{sec:Tail_Effect} and increasing further the quickness of our model.

\clearpage
\clearpage
\begin{table}
    \centering
    \begin{adjustbox}{angle=90}

%\caption{Enumeration Vuza canons.\label{enumvuza}}
			\tiny	\begin{tabular}{||c|@{\hspace{0.45\tabcolsep}} c@{\hspace{0.40\tabcolsep}}|@{\hspace{0.40\tabcolsep}} c @{\hspace{0.40\tabcolsep}}| c@{\hspace{0.40\tabcolsep}}|@{\hspace{0.45\tabcolsep}} c @{\hspace{0.40\tabcolsep}}|@{\hspace{0.45\tabcolsep}}c@{\hspace{0.45\tabcolsep}}|@{\hspace{0.45\tabcolsep}}c@{\hspace{0.45\tabcolsep}}|@{\hspace{0.45\tabcolsep}}c@{\hspace{0.45\tabcolsep}}|@{\hspace{0.45\tabcolsep}}c @{\hspace{0.45\tabcolsep}}||} 
				
			\hline
			$ n $&$ R_{A} $&$ R_{B} $& $n^{\circ}$ of $ A $'s& $n^\circ$ of $ B $'s & CSA $A$ & FP $A$ & CSA $B$& FP $B$\\ [0.5ex] 
			\hline\hline
			72 &$\{2,8,9,18,72\}$&$ \{3,4,6,12,24,36\} $& 6 (2)  &3 (1) &0.10 & 1.59 & 0.02 & 0.33\\
			\hline\hline
			108 &$ \{3,4,12,27,108\} $&$ \{2,6,9,18,36,54\} $&252 (30)&3 (1) &7.84 & 896.06 & 0.03 & 0.72\\
			\hline\hline
			120 &$ \{2,5,8,10,15,30,40,120\} $&$\{3, 4, 6, 12, 20, 24, 60\} $ &18 (4)&8  (2) &0.27 & 24.16 & 0.07 & 2.13\\
			\hline
			120 &$ \{2, 3, 6, 8, 15, 24, 30, 120\} $&$ \{4,5,10,12,20,40,60\} $&20 (3)&16 (5) &0.14 & 10.92 & 0.15 & 3.30\\
			\hline\hline
			144 &$ \{2,8,9,16,18,72,144\} $&$ \{3,4,6,12,24,36,48\} $&36 (10)&6 (1) &2.93 & 82.53  & 0.06 & 3.77\\
			%\hline
			%144 &$ \{2, 4, 9, 16, 18, 36, 144\} $&$ \{3,6,8,12,24,48,72\} $&8640 (591)&3 (1) &&&&\\
			\hline
			144 &$ \{4, 9, 16, 18, 36, 144\} $&$ \{2, 3, 6, 8, 12,18, 24, 48, 72\} $&6 (2)&12 (9) &0.10 & 7.13 & 1.71 & 66.27 \\
			&\color{lightgray}$ \{4, 9, 16, 18, 36, 144\} $&$ \{2, 3, 6, 8, 12, 24, 48, 72\} $&\color{lightgray}6 (2)&312 (1)&&&&\\
			\hline
			144&$ \{2,9,16,18,36,144\} $&$ \{3, 4, 6, 8, 12, 24, 36,48, 72\} $&12 (2)&6 (1)&0.11 & 12.13 & 1.08 &  33.39\\
			 & $\{2,9,16,18,144\} $&\color{lightgray}$ \{3, 4, 6, 8, 12, 24, 36,48, 72\} $&48 (7)&\color{lightgray}6 (1)&0.83&67.91&&\\
			 &\color{lightgray}$ \{2,9,16,18,36,144\} $&$ \{3, 4, 6, 8, 12, 24, 48, 72\} $&\color{lightgray}12 (2)&156 (9)&   &  & 1.71 & 74.78\\
			\hline\hline
			168 &$ \{2, 7, 8, 14, 21, 42, 56, 168\} $&$ \{3, 4, 6, 12, 24, 28, 84\} $&54 (8)&16 (3)& 17.61 & 461.53 & 0.13 & 7.91\\
			\hline
			168 &$ \{2, 3, 6, 8, 21, 24, 42, 168\}$&$ \{4, 7, 12, 14, 28, 56, 84\} $&42 (4)&104 (15)&0.91 & 46.11 & 1.94 & 35.36\\
			\hline\hline
			180 &$ \{3, 4, 5, 12, 15, 20, 45, 60, 180\} $&$ \{2, 6, 9, 10, 18, 30, 36, 90\} $&2052 (136)&8  (2)&1422.09  & >3600 & 0.25 & 1243.06\\
			\hline
			180 &$ \{2,5,9,10,18,20,45,90,180\} $&$ \{3, 4, 6, 12, 15, 30, 36, 60\} $&96 (12)&6 (1)& 48.04 & 900.75 & 0.11 & 8.22\\
			\hline
			180 &$ \{3, 4, 9, 12, 36, 45, 180\} $&$ \{2, 5, 6, 10, 15, 18, 20, 30, 60, 90\} $&1800 (171)&16 (5)& 492.18  & >3600 & 0.18  & 7.51\\%ricordarsi di controllare se sono effettivamente 1800 ritmi a meno di traslazione
			\hline
			180 &$ \{2,4,9,18,20,36,180\} $&$ \{3,5,6,10,12,15,30,45,60,90\} $&120 (18)& 9 (2)& 8.82& 280.72 & 0.29 & 14.34\\
			\hline
			
	\end{tabular}

\end{adjustbox}
    \small\caption{Comparison of Runtimes (in seconds) of the Cutting Sequential Algorithm (CSA) and the Fill-Out-Procedure (FP).}
    \label{table_times}
\end{table}

\small
\bibliographystyle{plain}      
\bibliography{biblio}  

\end{document}